\documentclass[11pt]{article}
\font\realfont=msbm10 at 11pt
\font\realfonta=msbm10 at 9pt
\usepackage{latexsym}
\newcommand{\bq}{\begin{equation}} 
\newcommand{\eq}{\end{equation}} 
\newcommand{\bqali}{\begin{eqnarray}}
\newcommand{\eqali}{\end{eqnarray}} 
\newcommand{\nein}{\noindent}
\newcommand{\real}{{\hbox{\realfont R}}}
\newcommand{\reala}{{\hbox{\realfonta R}}}

\title{T-duality and HKT manifolds}
\author{A. Opfermann\footnote{e-mail:ao200@damtp.cam.ac.uk}\\
\normalsize
\em Department of Applied Mathematics and Theoretical Physics, \\
\normalsize
\em University of Cambridge, \\
\normalsize
\em Silver Street, \\
\normalsize
\em Cambridge, CB3 9EW, UK}

\begin{document}

\maketitle

\begin{abstract}
We examine conformal rescaling and T-duality in the context of
four-dimensional HKT
geometries. The closure of the torsion forces the conformal factor 
to satisfy a
modified harmonic equation. Because of this equation the conformal factors form
non-commutative groups acting on the HKT geometries.
Using  conformal rescalings and
T-duality  transformations we generate from flat space new families of 
HKT geometries with tri-holomorphic Killing vectors.
We also find ultraviolet-finite (4,0) supersymmetric sigma models
which are not conformally invariant.

\bigskip

\noindent DAMTP R-97/44

\medskip

\noindent hep-th/9709048
\end{abstract}

\vfill


\section{Introduction}

It is well known that supersymmetry and the target space geometry of
sigma models are intimately related.  In particular, 
Alvarez-Gaume and
Freedman \cite{1}
 showed some time ago that (4,4) supersymmetry in two dimensional
sigma models
restricts the target space geometry to be hyper-K\"{a}hler (HK). 
More recently an
extension of the HK geometry, known as hyper-K\"{a}hler with torsion
(HKT) \cite{2aa}, 
was found in (4,0) supersymmetric sigma models
with Wess-Zumino coupling \cite{2,2a,2b,2c,2d}. 
While the former geometry is defined with respect to the 
Levi-Civita connection, the latter is defined 
with respect to a 
connection with torsion, 
the torsion $H$ being a closed three-form. In the limit
of vanishing torsion one recovers from an 
HKT geometry an HK
geometry. 

Apart from  their application to non-linear sigma models HKT
geometries were recently found as moduli spaces for a certain class of
black
holes \cite{3}. They also appear as the target spaces of
a bound state  of a D-string and D-five-branes in type I
string theory \cite{4}. Moreover, in \cite{2aa} it was shown that 
there is  a twistor space description associated to HKT geometries,
resembling the ordinary twistor space for HK geometries \cite{twist}.

The algebraic and differential constraints on
the couplings of an HKT sigma model target manifold, the metric $g$ and the
locally defined two-form $b$, are much better understood in four
dimensions than in
higher dimensions, and a number
of explicit examples of four-dimensional HKT geometries 
are known:  In \cite{5} it was shown,
that there
is a one-to-one correspondence between self-dual Einstein-Weyl spaces
and  HKT geometries. Moreover in \cite{8} it was shown, that  in the case
of a four-dimensional HKT geometry with a tri-holomorphic Killing vector
the associated three-space is a 'special' Einstein-Weyl space. The authors 
of \cite{6} have
used harmonic superspace methods to find generalizations of the
Eguchi-Hanson and Taub-NUT metrices to include torsion.
 Furthermore, it was shown in  \cite{7}
that there exists 
for every HK geometry with metric $g$ an associated family of HKT
geometries with metric
$\tilde{g}=U g$ and  torsion $H=-{}^{*_g} dU$, provided that U is harmonic with
respect to $g$. We generalize this observation, finding  a
family of conformally rescaled HKT geometries for every HKT geometry,
where the conformal factor is in general \textit{not} harmonic but rather
satisfies a modified harmonic equation.

We focus our attention on  HKT
ma\-ni\-folds, admitting at least one tri-holomorphic Killing vector;
furthermore we only consider four-dimensional HKT geometries. A
tri-holomorphic  Killing vector 
preserves all three complex structures. In addition, 
we always require an HKT geometry
with a tri-holomorphic Killing vector to have a 
torsion which is also  invariant under the transformations generated by this
Killing vector. Under these conditions one can T-dualize an HKT
geometry along the Killing direction to find a new HKT geometry \cite{9a}.

The torsion $H$ can be expressed through
 a one-form $u$ and a function $u_0$.
 For  $u_0=0$ the most general HKT geometry, which
admits a tri-holomorphic Killing vector, 
is  the conformal rescaling of the Gibbons-Hawking metric \cite{10},
 whereas for $u=0$ it is the geometry associated with monopoles on 
 the three-sphere \cite{11}.
We generate both of these metrices from flat
space by a series of consecutive T-duality
transformations and conformal re\-sca\-lings. Furthermore we find some 
new HKT geometries, which we  give in equations
 (\ref{ds4}), (\ref{ds8}), and (\ref{ds10}). 
  
The organisation of the paper is as follows: In section 2, the
HKT geometry is briefly reviewed, and
we focus on the case of  
HKT geometries with a
tri-holomorphic Killing vector. In section 3, we present the conformal
rescaling of an HKT geometry and show that the conformal factors form
non-commutive groups. In particular, we 
find the transformation rules for an
HKT geometry with a tri-holomorphic Killing vector 
under conformal rescaling and
T-duality. In section 4, starting from
ordinary flat space we apply successively conformal rescaling and
T-duality generating explicite examples of HKT manifolds. 
In section 5, we give some concluding remarks on the relation of
conformal invariance versus finiteness in (4,0) supersymmetric sigma models.

\section{The HKT geometry}
 
A manifold $M$ with a Riemannian metric g has an HKT
structure if it fulfils the following requirements \cite{2aa}: 

\nein (i) There is
a triplet of complex structures $\{ I_r;r=1,2,3\}$, 
obeying the quaternionic algebra
\bq
I_r I_s=-\delta_{rs}+\epsilon_{rst}I_{t} \ .
\label{quat}
\eq

\nein (ii) The complex structures are covariently constant

\bq
\nabla^{(+)}_{\mu} {(I_r)_{\nu}}^{\rho} = 0
\label{const}
\eq

\nein with respect to the connection with torsion

\bq
{\Gamma^{(+)\mu}}_{\nu\rho}= {\Gamma^{\mu}}_{\nu\rho} + \frac{1}{2} 
{H^{\mu}}_{\nu\rho} \ ,
\label{conn}
\eq

\nein where $\Gamma$ is the Levi-Civita connection with respect to the
metric $g$. 

\nein (iii) The metric $g$ is Hermitian with respect to all complex structures
\bq
g_{\rho (\mu} {(I_r)_{\nu)}}^{\rho}=0 \ .
\label{herm}
\eq

In the classical theory the torsion is a closed three-form, whereas 
in the quantum theory and in particular in the context
of the anomaly cancellation mechanism, the (classical) torsion
receives corrections \cite{18,19} 
and the new torsion is not a closed three-form
anymore. Throughout this paper we take the torsion to be closed, which
allows the local definition of a  two-form $b$ as

\bq
H_{\mu\nu\rho}=3 \ \partial_{[\mu}b_{\nu\rho]} \ .
\label{twoform}
\eq

\nein In particular in four
dimensions the torsion three-form $H$ is dual to a one-form $v$ 

\bq
H_{\mu\nu\rho} = {\epsilon_{\mu\nu\rho}}^{\lambda}v_{\lambda} \ ,
\label{dual}
\eq
where $\epsilon_{\mu\nu\rho\lambda}$ is the totally anti-symmetric 
tensor with respect to the
metric $g$. We remark that the conditions (i)-(iii)  are precisely the
conditions necessary for the invariance of the action of a 
sigma model under (4,0) supersymmetry and the closure of the
(4,0) supersymmetry algebra.

Now we assume that the sigma model is invariant under some target
space transformations generated by a vector field $X$. Equivalently
$X$ is an Killing vector field, which leaves the torsion $H$ and
the complex structures $I_r$ invariant

\bq
\mathcal{L}_X I_r = 0, \ \ \mathcal{L}_X H = 0 \ .
\label{tri}
\eq

\nein We call Killing vectors, which satisfy these equations,
tri-holomorphic.  Next, 
we choose coordinates $\{x_0, x_i; i=1,2,3\}$ on the target
space $M$ adapted to the Killing vector field $X$, i.e.

\bq
X= \frac{\partial}{\partial x_0} \ .
\label{adopt}
\eq

\nein The metric on $M$ in this adopted coordinate system can be
written as

\bq
ds^2=V^{-1}(dt+w_idx^i)^2 + V \gamma_{ij} dx^i dx^j \ ,
\label{metric}
\eq

\nein  where $\gamma$ is the three-dimensional metric on the manifold of
Killing vector trajectories, $M_3$, $V$ is a function and $w$ is a
one-form on
$M_3$. Following the notations of \cite{8} we write the torsion $H$ in
terms of a function  $u_0$ and a one-form $u$ on $M$ as

\bqali
H_{0ij} & = & {\epsilon_{ij}}^k u_k \nonumber \\
H_{ijk} & = & {\epsilon}_{ijk}(-V u_0 + w \cdot u) \ .
\label{defH}
\eqali

\nein We  choose an orientation such that
$\epsilon_{0123}=  + \sqrt{g} = + V \epsilon_{123} = + V \sqrt{\gamma}$.
The closure of the
torsion $H$ implies that the divergence of the one-form $u$ vanishes

\bq
\nabla^i u_i=0 \ ,
\label{close}
\eq

\nein where $\nabla_i$ is the Levi-Cevita connection
with respect to the three-metric $\gamma$.

Hence an  HKT geometry admitting a tri-holomorphic
Killing vector
is completely specified by a  
three-metric $\gamma$, and the tensors $V,w,u_0$
and $u$ on $M_3$, which fulfil the following restrictions \cite{8}:

\nein (i)  $V$ and $w$ satisfy a monopole-like equation

\bq
2 \partial_{[i}w_{j]} = {\epsilon_{ij}}^k(\partial_k V - V u_k)\ .
\label{mono1}
\eq

\nein (ii) $(\gamma_{ij},u_i)$ define a 
three-dimensional Einstein-Weyl geometry with a  Ricci tensor

\bq
R_{ij} - \nabla_{(i} u_{j)} + u_i u_j = \gamma_{ij} 
\Big(\frac{1}{2}u_0^2+|u|^2 \Big)\ .
\label{scalar}
\eq

\nein (iii) $(u_0, u_i)$ satisfy a monopole-like equation
analogous to the one for $(V,w)$

\bq
2 \partial_{[i}u_{j]} = { \epsilon_{ij}}^k( u_0 u_k - \partial_k u_0).
\label{mono2}
\eq
 
\nein An Einstein-Weyl space with the one-form $u$
restricted by (\ref{close}) and (\ref{mono2}) and the Ricci tensor given by
(\ref{scalar}) is called a \textit{special} Einstein-Weyl space. 

By taking the divergence of (\ref{mono1}) one finds that $V$ is a
solution of the modified harmonic equation

\bq 
\nabla^i \nabla_i V = u^i \nabla_i V,
\label{harmonic1}
\eq

\nein  which reduces to the ordinary harmonic
equation when
$u=0$. 
A similar equation follows from (\ref{mono2}) for $u_0$ as

\bq 
\nabla^i \nabla_i u_0 = u^i \nabla_i u_0 \ .
\label{harmonic2}
\eq


\section{Conformal rescaling and T-duality}

We start with the discussion of the conformal rescaling of a
general four-dimensional HKT geometry. Let $(g^1,H^1)$ be an HKT
geometry, then there is an HKT 
geometry $(g^2,H^2)$ given by 

\begin{eqnarray}
g^2 & = & U g^1\label{res1}\\
H^2 & = &-{}^{*_1} dU + U H^1,
\label{res1a}
\end{eqnarray}

\nein provided that the non-zero conformal factor $U$ satisfies

\bq 
\Box_1 U = (g^1)^{\mu\nu} (v_1)_{\mu} \partial_{\nu}U \ ,
\label{harmonicU}
\eq

\nein  where the Laplace operator $\Box_1$ is  taken 
with respect to the metric $g_1$. Note the factor of $U$ in front of $H^1$
 in (\ref{res1a}). This factor arises because  one of the indeces
 of $H$ is lowered in the first and second geometry
with different metrices
to find the torsion as a three-form. This factor is also responsible
for the fact that the conformal factor $U$ satisfies in general the modified
harmonic equation (\ref{harmonicU}) rather than a harmonic equation 
as it was claimed in \cite{8}.

The modified harmonic equation for $U$ is of such a form that the
conformal factors, to be precise the pairs 
$(U^{\alpha},H^{\alpha}; \alpha =1,2,3,...)$, form a
group that act on the space of the conformal 
families of  HKT geometries. In fact, the operation of conformal
rescaling closes separately on each conformal family
of HKT geometries. 
The group properties for the metrices $g^{\alpha}$ can be easily checked to be

\bqali
g^3  =  U^2 g^2 = U^2 \Big( U^1 g^1 \Big) = \Big( U^2 U^1 \Big) g^1 \hfill
& & \mathrm{closedness} \nonumber \\
g^4  =  \Big( U^3 U^2 \Big) U^1 g^1 = U^3 \Big( U^2 U^1 \Big) g^1 \hfill
& & \mathrm{associativity} \nonumber \\
 g^2  =  U^1 g^1 \ \Leftrightarrow \ \exists \ U^0=1 
\ \mathrm{s.t.} \  g^2  = \Big(U^0 U^1\Big) g^1  & & 
\mathrm{existence \ of \ unity} \nonumber \\
 \ g^2  =  U^1 g^1\ \Leftrightarrow \ \exists \ \Big(U^1\Big)^{-1} 
\ \mathrm{s.t.} \ 
g^1 = \Big(U^1\Big)^{-1} g^2 & & \mathrm{existence \ of \ inverse} 
\nonumber \ ,
\label{group}
\eqali

\nein where the torsions $H^{\alpha}$ 
are related to the metrices $g^{\alpha}$ as in
(\ref{res1a}). The group properties for the torsion and the conformal
factor are similar to the ones for the metric factors;
 for example for the closure of the
group one can show: (i) The torsion $H^3$ defined by 

\bq
H^3  = -{}^{*_2} dU^2 + U^2 H^2 \ ,
\label{closure1}
\eq

\nein can be rewritten as 

\bq
H^3  = -{}^{*_1} d(U^2 U^1) + (U^2 U^1) H^1 \ ,
\label{closure2}
\eq

\nein provided that the torsion $H^2$ is given as in (\ref{res1a}). (ii) The
 conformal factor
 $(U^2 U^1)$ is a harmonic function with respect to the metric $g^1$ if the
 conformal factors $U^1$ and $U^2$ are harmonic functions 
with respect to the metrices
 $g^1$ and $g^2$, respectively. Similarly one can proof the remaining
 group properties for the torsion and the conformal factor.
Note, that the group is not commutative because in 
  
\bq
g^3  =  U^2 \Big( U^1 g^1 \Big) = U^1 \Big( U^2 g^1 \Big) \nonumber
\label{groupa}
\eq

\nein $g=U^2 g^1$ is in general not an HKT metric.

Let us assume that the HKT geometry admits a tri-holomorphic Killing
vector. We  examine in detail, how such a geometry 
 transforms under conformal rescaling and T-duality. 
Let us first consider conformal rescaling of a tri-holomorphic HKT
geometry.

 Starting with an HKT geometry
$( (\gamma^1)_{ij}, V^1, (w^1)_i,(u^1)_0, (u^1)_i)$
we rescale the metric $g^1$ by a conformal factor $U$ to find 

\bq
(\gamma^2)_{ij}=U^2 (\gamma^1)_{ij}, \ V^2 = U^{-1} V^1, \
(w^2)_i=(w^1)_i \ .
\label{tricon}
\eq

\nein For the second geometry to be again tri-holomorphic, $U$ is
 restricted $U=U(x^i)$ and the dual torsion follows as

\bq
(u^2)_i=(u^1)_i-\partial_{i}(\ln U), \  (u^2)_0= U^{-1} (u^1)_0 \ .
\label{tricona}
\eq

\nein Equation (\ref{harmonicU}) 
reduces to the
three-dimensional modified harmonic equation

\bq
\nabla^i \nabla_i U = u^i \nabla_i U \ .
\label{triharmonicU}
\eq

Note, that for $u_0=0$ equation (\ref{mono2}) implies that $u_i$ is closed,
$\partial_{[i}u_{j]}=0$. Thus one can find locally a
function $f$ on $M$ such that
 $u_i= \partial_i f$. Then it follows from
(\ref{tricona}) that this geometry is conformally related to an HK
geometry. Since the most general HK geometry with a tri-holomorphic
Killing vector is the Gibbons-Hawking metric, an 
HKT geometry with $u_0=0$ is necessarily 
the conformal rescaling of  the Gibbons-Hawking metric.

The modified harmonic equation for $U$ (\ref{triharmonicU}) 
is similar to the equations 
(\ref{harmonic1}),  and (\ref{harmonic2}) for $V$, and $U_0$, 
respectively. To explain the fact that $U$ and $V$ satisfy similar
equations we remark that
in every conformal family of HKT geometries there are those for which
$V$ is a constant. ($V=const$ 
is a special solution of (\ref{harmonic1}).) 
 To find all the other metrices in the same conformal family
with non-constant $V$ we can conformally rescale the constant ones,
i.e. we have to solve the modified harmonic equation for $U$  
(\ref{triharmonicU}). Thus solving (\ref{triharmonicU}) for $U$ is equivalent
to solving in general (\ref{harmonic1}) for $V$.
The fact that $V$ and $u_0$ satisfy similar
equations enables one to find for every three-dimensional special Einstein-Weyl
space an associated 'minimal' four-metric  

\bq
V=-\frac{u_0}{c}, \ w = \frac{u}{c} 
\label{minimal}
\eq

\nein for any  constant $c$. In fact, every minimal HKT geometry is conformally
related to an HK geometry  \cite{8}.

We now examine the transformation of 
HKT manifolds, which admit a tri-holomorphic Killing vector, under T-duality.
Let us start again with an initial HKT geometry admitting a tri-holomorphic
Killing vector. Then T-dualizing this geometry leads to   \cite{9a}

\bq
(\gamma^3)_{ij}= (V^1)^2 \ (\gamma^1)_{ij}, \ V^3 =  (V^1)^{-1}, \
(w^3)_i=(b^1)_{0i} \ ,
\label{metric3}
\eq

\nein where $(b^1)_{0i}$ is a solution of the equation

\bq
2\partial_{[i}(b^1)_{j]0}={(\epsilon^1)_{ij}}^k(u^1)_k \ .
\label{defb}
\eq

\nein There always exists such a $(b^1)_{0i}$ because the integrability
condition for (\ref{defb}) is just (\ref{close}).
The new torsion follows as 

\bq 
(u^3)_i=(u^1)_1-\partial_{i}(\ln V^1), \  (u^3)_0= V^{-1} (u^1)_0 \ .
\label{metric3a}
\eq

One can explicitly show that  the conformal
rescaling (\ref{tricon}) and T-dua\-lization (\ref{metric3}) 
of an initial   geometry give rise to further HKT geometries,  iff
  the initial geometry  itself is HKT, i.e. equations 
(\ref{close}) -
(\ref{mono2}) are invariant under conformal rescaling and T-duality,
as expected on general grounds.


\section{HKT metrices from flat space}

We construct explicit examples of  HKT manifolds
with a tri-holomorphic Killing vector using a string of consecutive
T-duality transformations and conformal rescalings.
Let us start with flat space as a trivial example of an HKT geometry
\footnote{ We can  equally well take the space $S^1 \times 
\reala^3$ 
as a starting point and the periodic identification of the Killing
direction goes through all T-dualizations and conformal rescalings.}

\bqali
g^1 & = & (dx_0)^2 + d x \cdot d x \nonumber \\ 
v^1 &  = &  0 \ .
\label{ds1}
\eqali

\nein  Now conformally rescale the metric (\ref{ds1})
as in (\ref{tricon})
 with $U^1=U^1(x^i)$ satisfying $\nabla^i \nabla_i U=0$ to
find

\bqali
g^2 &=&  U^1 \Big( (dx_0)^2 + d x \cdot d x \Big) \nonumber \\ 
(u^2)_0  &=&   0 \ \ , (u^2)_i= - \partial_i(\ln U^1) \ .
\label{ds2}
\eqali

\nein A one-form $w^1$ on $M_3$, defined  as

\bq
2 \partial_{[i} (w^1)_{j]} = {(\epsilon^1)_{ij}}^k \partial_{k} U^1 \ ,
\label{w1}
\eq

\nein  always exists because the integrability condition for
(\ref{w1}) is that $U$ is harmonic. In a  next
step we T-dualize the geometry (\ref{ds2}) to find 

\bqali
g^3 &  = & (U^1)^{-1}(dx_0 + (w^1)_i dx^i)^2 + 
U^1 d x \cdot d x \nonumber \\ 
v^3 & = & 0 \ ,
\label{ds3}
\eqali

\nein This is the well known HK metric of Gibbons and Hawking
 \cite{10}.  Let us write
the flat metric on $M_3$ in spherical polar coordinates

\bq
\gamma^3= dr^2 + r^2 ( d\phi^2 + \sin^2 \! \phi \, d\psi^2) \ 
\label{gamma3}
\eq

\nein and take $U^1$ as the one-centre Greens function on $M_3$

\bq
U^1=c_0 + \frac{c_1}{r} \ 
\label{defU1}
\eq

\nein with $c_0$ and $c_1$ as non-negative constants. Then it can be arranged
that 

\bq
w^1= -c_1 \cos \! \phi \, d\psi \ .
\label{defw1}
\eq

The Gibbons-Hawking metric (\ref{ds3}) can be further conformally
rescaled 
 by $U^2$, which satisfies a similar harmonic
equation on the flat three-space 
as  $U^1$. In particular, we  can take $U^2$ and $w^2$ to be of the same
form as $U^1$ and $w^1$

\bqali
U^2&=&c_2 + \frac{c_3}{r} \nonumber \\
w^2&= &-c_3 \cos \! \phi \,d\psi \ 
\label{defU2}\
\eqali

\nein with $c_2$ and $c_3$ as non-negative constants. Changing variables
from $r$ to $z=\ln r$, $-\infty < z < + \infty$, the conformal
rescaling leads to 

\bqali
g^4&=& \frac{c_3 + c_2 e^z}{c_1 + c_0 e^z}\Big(dx_0 - c_1 \cos \! \phi
\, d\psi\Big)^2 + \nonumber \\
&& \frac{c_1 + c_0 e^z}{c_3 + c_2 e^z} \Big[(c_3 + c_2 e^z)^2 (d z^2 +
d\phi^2 + \sin^2 \! \phi \, d\psi^2)\Big] \nonumber \\
(u^4)_0 & = &(u^4)_\phi=(u^4)_\psi =  0 , \ 
(u^4)_z=\frac{c_3}{c_3 + c_2 e^z}\ .
\label{ds4}
\eqali

\nein In the generic case of
all four constants $\{c_1, c_2, c_3, c_4\}$ to be
positive, the metric factor $V^4$ is regular but the
three-metric is singular at $z = + \infty$, although  
the  singularity is in an infinite proper distance.
For $c_0=c_2=0$, $c_1=\lambda$ and $c_3=1$ the
metric and the dual torsion (\ref{ds4}) become

\bqali
g^{5}&=& {\lambda}^{-1}(dx_0 -\lambda \, \cos \, \phi\, d\psi)^2 + \lambda
 (d z^2 + d\phi^2 + \sin^2 \phi \, d\psi^2) 
\nonumber \\
(u^{5})_0 & = &(u^{5})_\phi=(u^{5})_\psi =  0 , \ (u^{5})_z = 1 \ .
\label{ds5}
\eqali

\nein This geometry  has a constant $V^5=\lambda$ and its 
three-geometry is $\real \times S^2$. Though $v^{5}$ is constant,
$H^{5}$ is not because $\epsilon_{ijk}$ is not constant on $\real
 \times S^2$.
The coordinate $z$ can be periodically identified to give the compact
version of the three-space, $S^1 \times S^2$. Both three-geometries as
well as their associated four-geometries are \textit{non-singular}.
The effect of a further T-dualization of the
geometry (\ref{ds4}) is just to interchange  the constants $(c_0,c_1)$ with 
$(c_2,c_3)$. Therefore the geometry (\ref{ds5}) with $\lambda=1$ 
is T-self-dual. Thus we cannot generate from (\ref{ds4}) 
any other HKT manifolds by further 
conformal rescalings or 
 T-dualizations, i.e we are confined to one particular conformal family.  

Let us start again with the flat  four-space  (\ref{ds1}) but 
rewritten as

\bqali
g^6 & = & dr^2 + r^2 d\, \tilde{\Omega}_3^2 \nonumber \\ 
v^6 &  = & 0 \ ,
\label{ds6}
\eqali

\nein where $d\, \tilde{\Omega}_3^2$ is the volume element on the unit three
sphere, $S^3$. We perform a first conformal rescaling of (\ref{ds6})
by a harmonic function $U^3$ which  depends on $r$ only,

\bq
U^3=\frac{R^2}{r^2} \ 
\label{defU3}
\eq

\nein with $R$ a poitive constant. Thus the rescaled geometry is

\bqali
g^7& =&dx_0^2+d\, \Omega_3^2 \nonumber \\
(u^7)_0 &=& \frac{2}{R}, \ (u^7)_i=0 \ ,
\label{ds7}
\eqali

\nein where $x_0= R \ln r$, and $d\, \Omega_3^2$ is the volume
element on the three-sphere with radius $R$, \  
$d\, \Omega_3^2 = R^2 d\, \tilde{\Omega}_3^2$. Note that this
conformal rescaling is not of the form $U=U(x^i)$ as all the others are.
With the former
type of conformal rescaling we have generated a non-zero 
$u_0$, whereas with the
latter we  cannot make $u_0$ non-zero. In fact, the geometry (\ref{ds7})
is a member of a
new conformal family. Even though the conformal
rescaling does not preserve a tri-holomorphic Killing vector, after the 
 coordinate change from $r$ to $x_0$, the geometry is cast in the form
of an HKT geometry with a  tri-holomorphic Killing vector. Further
rescaling the metric (\ref{ds7}) by $U^4$ gives

\bqali
g^8& =&U^4(dx_0^2+d\, \Omega_3^2) \nonumber \\
(u^8)_0 &=& \frac{2}{U^4 R}, \ (u^8)_i= -\partial(\ln U^4) \ ,
\label{ds8}
\eqali

\nein where $U^4$ is chosen to be a harmonic function 
on the three-sphere. Define
$w^4$ by

\bq
2 \partial_{[i} (w^4)_{j]} = {(\epsilon^4)_{ij}}^k \partial_{k} U^4 \ 
\label{w4}
\eq

\nein with $\epsilon^4$  taken with respect to $d\, \Omega_3^2$.
T-dualization of   (\ref{ds8}) leads to

\bqali
g^9 &  = & (U^4)^{-1}(dx_0 + (w^4)_i dx^i)^2 + 
U^4  d \Omega_3^2 \nonumber \\ 
(u^9)_0&=&\frac{2}{R}, \ (u^9)_i =  0 \ .
\label{ds9}
\eqali

\nein This is the unique HKT geometry with
$u=0$. This geometry was given in \cite{11} 
with $R=\lambda^{-1}$. The three-geometry of (\ref{ds9}) 
is that of a
three-sphere of constant radius $R$. The global behaviour of
(\ref{ds9}) was studied in \cite{11} for the case of a
geometry, which is $u(2)$-invariant. It was 
found that all geometries of this kind are singular except the WZW model with
target space $SU(2)\times
U(1)$, which has a constant $V$ and a vanishing $w$. It seems
likely that all the remaining metrices of the type (\ref{ds9}) are
also  singular.
Further rescaling by a conformal
factor $U^5$, which is again a harmonic function on $S^3$, results in  

\bqali
g^{10} &  = & \frac{U^5}{U^4} \Big(dx_0 + (w^4)_i dx^i\Big)^2 + 
 \frac{U^4}{U^5}\Big[(U^5)^2  d \Omega_3^2\Big] \nonumber \\ 
(u^{10})_0&=&\frac{2}{U^5 R}, \ (u^{10})_i =  -\partial_i (\ln U^5) \ .
\label{ds10}
\eqali

We have checked that  the singularities of the metric
(\ref{ds9}), which is also $u_2$-invariant, cannot  be resolved by such
a  conformal
rescaling. Note, that the effect of T-duality on (\ref{ds10}) is to 
interchange the pairs $(U^4, w^4)$ with  $(U^5, w^5)$ which
resembles the behaviour of  (\ref{ds5}) under T-duality. 
Thus again applying our techniques  to (\ref{ds10}) does not yield any
 new geometries.


\section{Concluding remarks}

It has been shown in  \cite{2d, finite1} by a power counting argument 
that (4,0) supersymmetric sigma models are ultraviolet finite. 
The condition for one-loop ultraviolet finiteness is \cite{finite2, finite3}

\bq
R_{ij}^{(+)}= \nabla_{(i} V_{j)} + \frac{1}{2} T_{ij}{}^k V_k +
\partial_{[i} \lambda_{j]} \ ,
\label{UV}
\eq

\nein where $R^{(+)}$ is the curvature associated with $\Gamma^{(+)}$,
 and $V_i$ and $\lambda_i$ are one-forms on
$M$. In comparison, the condition for one-loop conformal 
invariance is \cite{finite3}

\bq 
R_{ij}^{(+)}= \nabla_i \nabla_j \phi + \frac{1}{2} T_{ij}{}^k \nabla_k \phi \ 
\label{inv}
\eq

\nein for some function $\phi$ on $M$. The two conditions coincide
only if $V_i$ is a closed one-form, i.e. one-loop conformal invariance implies
one-loop ultraviolet finiteness  but not vice-versa. 

The Ricci-tensor for an HKT geometry is given in \cite{6} as

\bq
R_{ij}^{(+)}= \nabla^{(+)}_j v_i \ ,
\label{genR}
\eq

\nein where $v_i$ is the dual torsion (\ref{dual}). 
It follows that the condition for
one-loop ultraviolet finiteness (\ref{UV}) is fulfilled for any HKT
geometry with
$\lambda_i= -v_i$ and $V_i=v_i$, whereas the  condition for one-loop
conformal invariance (\ref{inv}) is fulfilled only for geometries
for which one can set $\nabla_i \phi = v_i$. Thus all HKT geometries are
one-loop ultraviolet finite, but only conformally rescaled HK
geometries are one-loop conformally invariant. Moreover in
\cite{polch} it was shown that for compact target spaces any
ultraviolet finite sigma model is conformally invariant. We remark
though that the (4,0) supersymmetric sigma model with target space the
HKT geometry (\ref{ds9}) is ultraviolet finite but not conformally
invariant. 

\bigskip
\nein {\bf Acknowledgments:} The author would like to thank G.
Papadopoulos for helpful discussions and advice and K. Sfetsos for
discussions about the last section. The author also
thanks EPSRC and the German National Foundation for studentships.



\end{document}